\documentclass[aps,prl,twocolumn,showpacs,superscriptaddress]{revtex4-1}
\usepackage{bm}
\usepackage{mathrsfs}
\usepackage{amsmath}
\usepackage{amssymb}
\usepackage{graphicx}
\usepackage{amsfonts}
\usepackage{amsthm}
\usepackage{color}
\usepackage{dcolumn}
\usepackage{txfonts}

\begin{document}

\title{Extracting the Central Charge of Conformal Field Theory by Central Spin Decoherence}
\author{Bo-Bo Wei}
\email{Corresponding author: bbweiphy@gmail.com}
\affiliation{School of Physics and Energy, Shenzhen University, Shenzhen 518060, China}

\begin{abstract}
Conformal invariance powerfully constrains the critical behavior of two-dimensional classical systems with short-range interactions and the critical theories in two-dimensions are parametrized by a dimensional number, termed central charge or conformal anomaly $c$.
However, experimental determination of the central charge of a conformal field theory has not been done before. Here we propose to extract the central charge of the conformal field theory corresponding to a critical point of a two-dimensional lattice models from the quantum decoherence measurement of a probe spin which is coupled to the two-dimensional lattice models.  Conformal invariance predicts that the leading finite-size correction to the free energy for a two-dimensional system at a conformal invariant critical point is linearly related to the conformal anomaly for various boundary conditions. A recent discovery of thermodynamic holography allows us to obtain the free energy of many-body system from central spin decoherence measurement. Thus the central charge of the conformal field theory could be extracted from central spin decoherence measurement. We have applied the method to the two-dimensional Ising model and extracted the central charge of two-dimensional Ising model with good precision. This work provides a useful approach to extracting the central charge of conformal field theory in various two-dimensional lattice systems.
\end{abstract}


\maketitle

\emph{Introduction.-}
Scale invariance is one of the most intriguing features of critical points of phase transitions in statistical mechanics \cite{Stanley1971,Cardy1996}. Local scale invariance, i.e., conformal invariance, at a critical
point have been demonstrated to be remarkably powerful, especially in two dimensions \cite{BPZ1984,conformalbook}. In two
dimensions the group of conformal transformations
is exceptionally large (infinite dimension) because
any analytic function mapping the complex plane to
itself is conformal. Universality classes of critical phenomena in two-dimensions is characterized by a single dimensionless
number $c$, the conformal anomaly or the value of the
central charge of the Virasoro algebra \cite{Virasoro1970}. lt was demonstrated that \cite{Shenker1984} unitarity constrains
the values of $c<1$ to be quantized, $c=1-6/[m(m+1)]$ with $m=3,4,5,\cdots$. For
such theories, the critical exponents are given by the
Kac formula \cite{Kac1979}.

Universality classes of critical phenomena in two-dimensions are classified by the central charge $c$ of corresponding conformal field theory \cite{conformalbook}. However, how to experimentally extract the central charge of the conformal field theory corresponding to a critical point of a two-dimensional lattice models has not been known before. Conformal invariance has been used to study finite-size effects in two-dimensional statistical systems \cite{Cardy1985,Affleck1985} and it was found that leading finite-size correction to the free energy for a two-dimensional system at a conformal invariant critical point is linearly related to the conformal anomaly for various boundary conditions \cite{Cardy1985,Affleck1985}. Thus experimental measurement of the free energy of a two-dimensional many-body system shall tell us the central charge of the conformal field theory.  A recent discovery of thermodynamic holography by the author and its collaborators allows us to obtain the free energy of many-body system from central spin decoherence measurement \cite{Wei2012}. Therefore we propose to extract the central charge of the conformal field theory corresponding to a critical point of a two-dimensional lattice models from the quantum decoherence measurement of a probe spin which is coupled to the two-dimensional lattice models.

\emph{Universal Term of the Free Energy at Critical Point in Two-Dimensions.-}
According to renormalization group theory, a many-body system at critical point is governed by a reduced
fixed-point Hamiltonian $H^*$  \cite{Cardy1996}. Under a uniform rescaling of the lengthes, the form
of the Hamiltonian is invariant. For systems with short-range interactions, the Hamiltonian remains at the fixed point
under conformal transformations, which corresponds to a non-uniform scaling of the lengthes and rotation. Transformations with a shear component (non-conformal)
indeed modify the Hamiltonian. The response of $H$ to such
an infinitesimal (non-conformal) transformation of the form $r^{\mu}\rightarrow r^{\mu}+\alpha^{\mu}(r)$,
\begin{eqnarray}
\delta H=-\frac{1}{2\pi}\int d^2zT_{\mu\nu}\partial^{\mu}\alpha^{\nu}.
\end{eqnarray}
This equation defines the stress tensor $T_{\mu\nu}$. Under a general conformal transformation $z\rightarrow z'$, the stress tensor transforms $T\rightarrow T'$ according to
\begin{eqnarray}\label{c1}
T(z)=T'(z')\Big(\frac{dz'}{dz}\Big)^2+\frac{c}{12}\{z',z\},
\end{eqnarray}
where $\{z',z\}$ is the Schwarzian derivative of the transformation and defined by
\begin{eqnarray}
\{z',z\}=\frac{d^3z'/dz^3}{dz'/dz}-\frac{3}{2}\Big(\frac{d^2z'/dz^2}{dz'/dz}\Big)^2.
\end{eqnarray}

Let us consider the logarithm transformation, $z'=w(z)=\frac{L}{2\pi}\ln z$. Under this transformation, the infinite complex plane parametrized by $z$ is transformed into the infinitely long strip of finite width $L$ with
periodic boundary condition (Figure 1). Due to the rotational invariance of the infinite complex plane geometry, we can assume that $\langle T\rangle_z=0$. As a consequence of Equation \eqref{c1}, one has
\begin{eqnarray}
\langle T'\rangle_{w}=\Big(\frac{2\pi}{L}\Big)^2\frac{c}{24}.
\end{eqnarray}
An infinitesimal deformation (which may not be conformal) of a domain $D$
induces a corresponding change of the free energy by \cite{Cardy1988}
\begin{eqnarray}
\delta \ln Z+\int_D\frac{d^2z}{2\pi}\langle T_{\mu\nu}\rangle\partial^{\mu}\alpha^{\nu}=0.
\end{eqnarray}
where $\langle T_{\mu\nu}\rangle$ is the expectation value of the stress tensor and the integral extends over the domain $D$. Applying this to a transverse dilatation of the infinitely long strip $\delta\Re z=\epsilon \Re z, \delta \Im z=0$, one have the free energy in a $L\times L'$ two-dimensional lattice \cite{Affleck1985,Cardy1985}
\begin{eqnarray}\label{centralA}
f=-\lim_{L\rightarrow\infty}\frac{\ln Z_{LL'}}{LL'}=A-\frac{\pi c}{6L'^2}.
\end{eqnarray}
Here the first term depends on the area and is non-universal. However the second term is universal and depends only on the central charge or conformal anomaly $c$. Equation \eqref{centralA} provides a direct access to the determination of central charge $c$ using finite-size scaling methods of the free energy at the critical point of a two-dimensional system with short-range interactions.

For rectangular lattices with periodic boundary conditions, the corrections of free energy to the strip geometry is \cite{Torus1986}
\begin{eqnarray}\label{centralB}
F=-\ln Z_{LL'}=AS-\frac{\pi c}{6}\Big(x+\frac{1}{x}\Big)+o(x^2).
\end{eqnarray}
Here $S=L\times L'$ is the area of the rectangular domain and $x\equiv L/L'$ is the aspect ratio of the two edges. As $x\rightarrow0$ or $\infty$, Equation \eqref{centralB} extends to Equation \eqref{centralA}. Now for system with $c>0$, if the area is fixed and the shape varied, the approximate $F$ is
maximal for a square $(x = 1)$. Thus there is a thermodynamic driving
force for elongation of the domain and may be understood due to an attraction of
the walls of the rectangle with a force per unit length inversely proportional to the
square of their separation. From this perspective, the tendency to elongation should be a general geometric effect,
not limited to domains of rectangular shape. To test these effects, experimental measurement of the free energy of a many-body system is desirable.

\begin{figure}
\begin{center}
\includegraphics[width=\columnwidth]{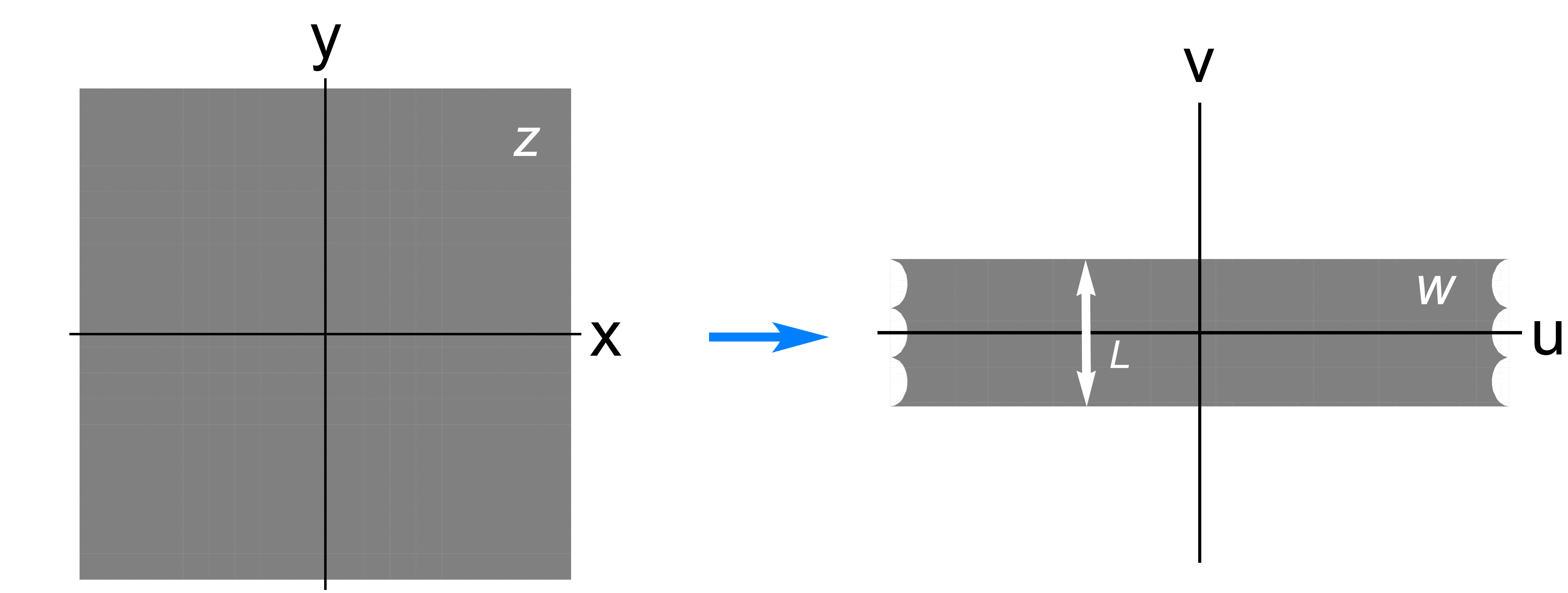}
\end{center}
\caption{(color online). Logarithmic transformation in two-dimensions, $w(z)=u+iv=\frac{L}{2\pi}\ln z$. which transforms the whole complex plane parametrized by $z$ into an infinitely long strip of finite width $L$ with a periodic boundary condition parametrized by $w=u+iv$.}
\label{fig:epsart1}
\end{figure}

\begin{figure}
\begin{center}
\includegraphics[width=\columnwidth]{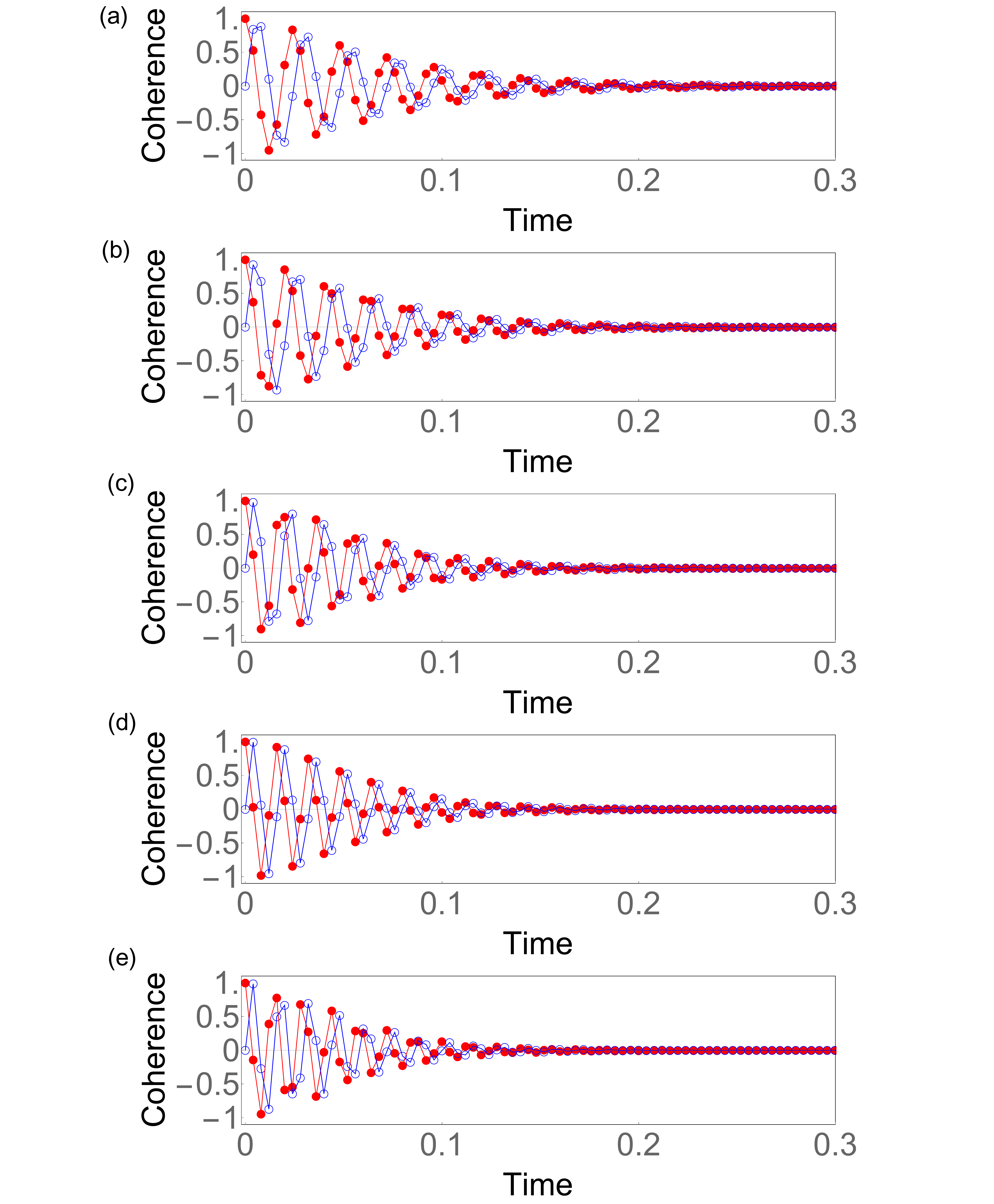}
\end{center}
\caption{(color online). Quantum decoherence of the probe spin which is coupled to the two-dimensional Ising model at inverse temperature $\beta=\beta_c$ and magnetic field $h=0.1J$ as a function of time for different sizes of rectangle. The red solid dot is the real part of the coherence $\langle S_x(t)\rangle$ and the blue circle is the imaginary part of the coherence $\langle S_y(t)\rangle$. (a). Spin decoherence for $6\times50$ rectangular lattices. (b). Spin decoherence for $7\times50$ rectangular lattices. (c). Spin decoherence for $8\times50$ rectangular lattices. (d). Spin decoherence for $9\times50$ rectangular lattices. (e). Spin decoherence for $10\times50$ rectangular lattices. }
\label{fig:epsart2}
\end{figure}

\emph{Free Energy and Central Spin Decoherence.-} To access the central charge of the conformal field theory, the key is to devise a free energy measurement in a many-body systems. Previous studies shows that the central spin decoherence and free energy are deeply connected \cite{Wei2012,Wei2014,Wei2015,Peng2015,LYExp2015,Wei2017}, which we briefly explain below.
Let us consider a general many-body system with Hamiltonian
\begin{equation}\label{ham}
H(\lambda)=H_0+\lambda H_1.
\end{equation}
where $H_0$ and $H_1$ are two competing Hamiltonian and $\lambda$ is a control parameter.  We use a probe spin-1/2 coupled to the many-body system (bath), with probe-bath interaction
$H_I=-\eta S_z\otimes H_1 $ and $\eta$ being a coupling constant and $S_z\equiv(|\uparrow\rangle\langle\uparrow|-|\downarrow\rangle\langle\downarrow|)/2$
being the Pauli matrix of the probe spin. If we initialize the probe spin in a superposition state as
$(|\uparrow\rangle+|\downarrow\rangle)/\sqrt{2}$ and the bath at inverse temperature
$\beta=1/T$ descried by $\rho_0=e^{-\beta H}/Z(\beta,\lambda)$ with $Z(\beta,\lambda)=\text{Tr}[e^{-\beta H(\lambda)}]$ being the partition function of the bath.  Then the quantum coherence of the probe spin, defined as $\langle S_+(t)\rangle\equiv\langle S_x\rangle+i\langle S_y\rangle$, has an intriguing form as \cite{Wei2012,Wei2014}
\begin{eqnarray}\label{central1}
\langle S_+(t)\rangle=\frac{Z(\beta,\lambda+it\eta/\beta)}{Z(\beta,\lambda)}.
\end{eqnarray}
The denominator in the above equation is nonzero for real control parameter $\lambda$. The numerator resembles the form of a partition
function but with a complex control parameter $\lambda+it\eta/\beta$. Equation\eqref{central1} establishes the relation between partition function with a complex parameter and the central spin decoherence, which leads to the first experimental observation of Lee-Yang zeros \cite{Peng2015,LYExp2015}.

The relation between free energy and the central spin decoherence are established by the thermodynamic holography \cite{Wei2015}, which states that the partition function of a finite many-body system is an analytic function of its control parameters and thus according to Cauchy theorem, the partition function in the whole complex plane are uniquely determined by its values along a closed contour in the complex plane of the physical parameter\cite{Wei2015}. Consider a rectangular closed contour where two vertical lines with real parts $\lambda_1$ and $\lambda_2$ respectively, and we assume $\lambda_1<\lambda_2$ and the two horizontal lines with real parts extends from $\lambda_1$ to $\lambda_2$ and with imaginary parts being $-\infty$ and $\infty$, respectively. The contributions from two horizontal lines vanishes and we then have
\begin{eqnarray}\label{central2}
Z(\beta,\lambda')&=&\int_{-\infty}^{\infty}\frac{d\lambda_I}{2\pi}\frac{Z(\lambda_2+i\lambda_I)}{\lambda_2+i\lambda_I-\lambda'}
-\int_{-\infty}^{\infty}\frac{d\lambda_I}{2\pi}\frac{Z(\lambda_1+i\lambda_I)}{\lambda_1+i\lambda_I-\lambda'}.\nonumber\\
\end{eqnarray}
Here $\lambda_1<\Re\lambda'<\lambda_2$. Combing with Equation \eqref{central1}, we have
\begin{eqnarray}\label{central3}
e^{-\beta F(\lambda')}&=&e^{-\beta F(\lambda_2)}\int_{-\infty}^{\infty}\frac{d\eta t}{2\pi}\frac{\langle S_+(\lambda_2,t)\rangle}{\beta\lambda_2+i\eta t-\beta\lambda'}\nonumber\\
&&-e^{-\beta F(\lambda_1)}\int_{-\infty}^{\infty}\frac{d\eta t}{2\pi}\frac{\langle S_+(\lambda_1,t)\rangle}{\beta\lambda_1+i\eta t-\beta\lambda'}.
\end{eqnarray}
Here $F=-\beta^{-1}\ln Z$ is the Helmholtz free energy. $\langle S_+(\lambda,t)\rangle$ is the spin coherence at time $t$ for the probe spin which is coupled to the bath with Hamiltonian $H(\lambda)$. In the case of no ambiguity, we suppress the bath parameter $\lambda$ in $\langle S_+(\lambda,t)\rangle$. Because $\langle S_+(\lambda,t)\rangle^*=\langle S_+(\lambda,-t)\rangle$, Equation \eqref{central3} implies that one can obtain the free energy at parameters $\lambda'$ with $\lambda_1<\Re\lambda'<\lambda_2$ by measuring the central spin coherence $\langle S_+(\lambda,t)\rangle$ at two bath parameters $\lambda_1$ and $\lambda_2$ for $t\in(0,\infty)$.

If the system has a symmetry such that $UH(\lambda)U^{\dagger}=H(-\lambda)$, this leads to $Z(\beta,\lambda)=Z(\beta,-\lambda)$. In such a case, we can choose, $\lambda_2=-\lambda_1=-\lambda$, then we have
\begin{eqnarray}\label{central4}
\frac{Z(\beta,\lambda')}{Z(\beta,\lambda)}&=&\int_{-\infty}^{\infty}\frac{d(\eta t)}{\pi}\langle S_+(\lambda,t)\rangle\Bigg[\frac{(\beta\lambda+i\eta t)}{(\beta\lambda+i\eta t)^2-(\beta\lambda')^2}\Bigg].
\end{eqnarray}
For classical lattice models, the central spin coherence $\langle S_+(\lambda,t)\rangle$ is a periodic function of time with period $T$, then Equation \eqref{central4} can be simplified further into,
\begin{eqnarray}\label{central5}
\frac{Z(\beta,\lambda')}{Z(\beta,\lambda)}&=&\int_{0}^{T}\frac{d(\eta t)}{2\pi}\langle S_+(\lambda,t)\rangle\Big[\frac{\sinh(\beta\lambda+i\eta t)}{\cosh(\beta\lambda+i\eta t)-\cosh(\beta\lambda')}\Big].\nonumber\\
\end{eqnarray}
Equation \eqref{central5} implies that one can obtain the free energy at any parameters from central spin coherence data within one period $t\in [0,T)$. Since the central spin decoherence is directly experimentally measurable \cite{scexp1,scexp2,scexp3,scexp4,scexp5}, we can extract the central charge of conformal field theory associated to a critical point in two-dimension from central spin decoherence measurement by combing Equation \eqref{central5} and Equation \eqref{centralA}.

\begin{figure}
\begin{center}
\includegraphics[width=\columnwidth]{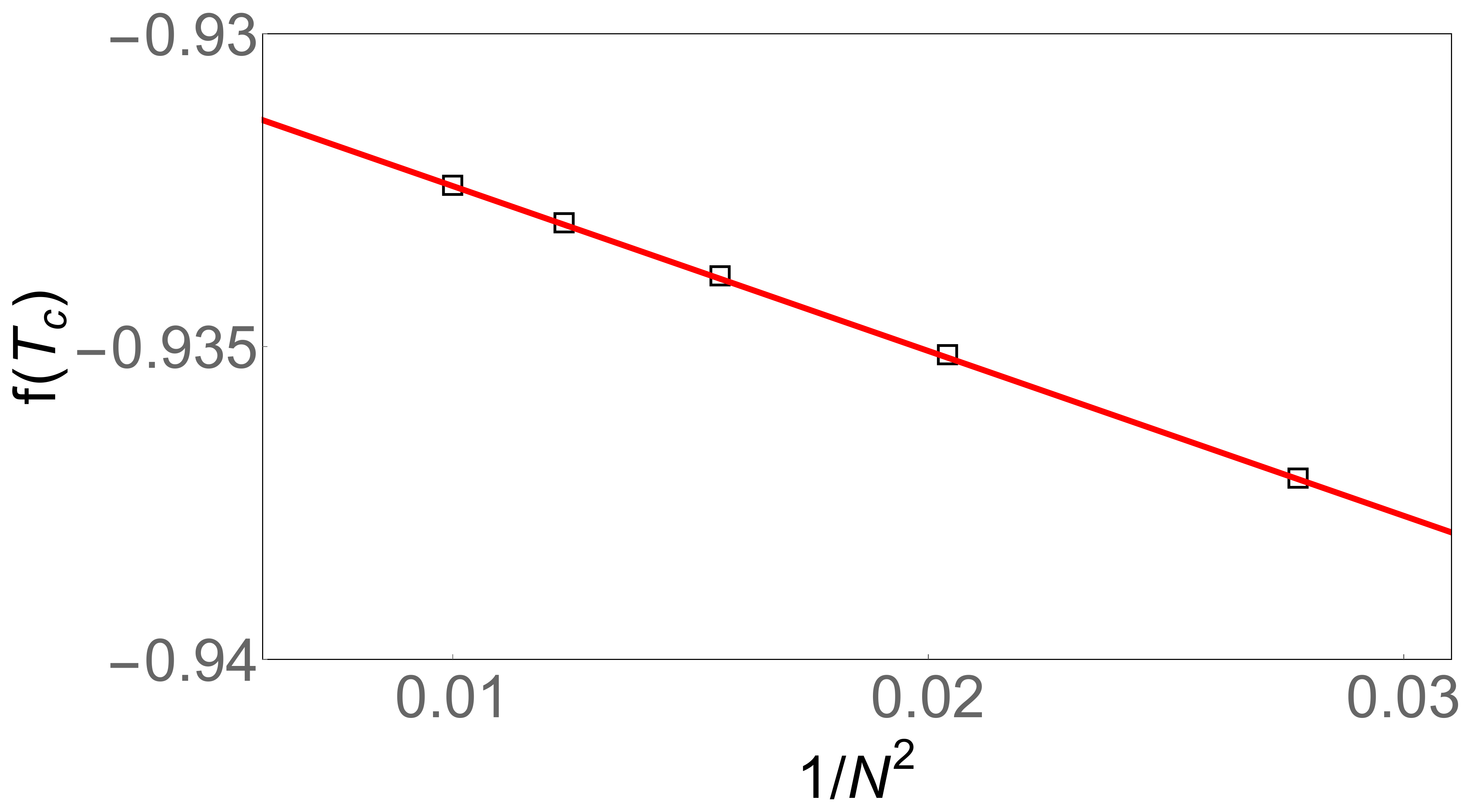}
\end{center}
\caption{(color online). Extracting the central charge of conformal field theory corresponding to the critical point of the two-dimensional Ising model. The squares denote the free energy reconstructed from the spin coherence data (Figure 2) by Equation \eqref{a}. }
\label{fig:epsart3}
\end{figure}

\emph{Study of Two-dimensional Ising Model.-}
To illustrate the above idea, we study the two-dimensional (2D) Ising model in a square lattice with $N$ rows and $M$ column with Hamiltonian
\begin{eqnarray}
H&=&-J\sum_{i=1}^N\sum_{j=1}^M(\sigma_{i,j}\sigma_{i+1,j}+\sigma_{i,j}\sigma_{i,j+1})-h\sum_{i=1}^N\sum_{j=1}^M\sigma_{i,j}.
\end{eqnarray}
Here $\sigma_{i,j}=\pm1$ is the spin in the site $(i,j)$ and $J$ is the ferromagnetic coupling constant between nearest-neighbor spins in the lattice and $h$ is the magnetic field experienced by all the spins and periodic boundary conditions are applied. The 2D Ising model at zero magnetic field $h=0$ has been exactly solved by Onsager in 1944 \cite{Onsager1944} and there is a finite temperature phase transition at $\beta_c=\ln(1+\sqrt{5})/2$ \cite{Onsager1944,Kramers1941}. For 2D Ising model under a finite magnetic field, there is no exact solution available but one can map the problem into 1D quantum Ising model with both longitudinal and transverse field by transfer matrix method \cite{Schultz1964,Wu1973}.

Although the two-dimensional Ising model in a finite magnetic field has not been exactly solvable up to now, we can obtain the partition function of the 2D Ising model at any finite magnetic field from spin decoherence measurement of the probe spin which is coupled to the 2D Ising model at zero magnetic field for time $t\in(0,T)$ from Equation \eqref{central5}.

The 2D Ising critical point occurs at zero magnetic field \cite{Onsager1944}. From Equation \eqref{central5}, the free energy at zero magnetic field is
\begin{eqnarray}\label{a}
\frac{Z(\beta,0)}{Z(\beta,h)}=\int_{0}^{T}\frac{\langle S_+(h,t)\rangle}{\tanh[(\beta h+i\eta t)/2]}\frac{d(\eta t)}{2\pi}.
\end{eqnarray}
This is the central formula to extract the free energy at the Ising critical point from the central spin decoherence data within one period. $Z(\beta,h)$ is the partition function away from the critical point and can be obtained from Viral expansion method \cite{Reichl2016}. An alternative method to get the partition
function of a finite system away from critical point is the cluster correlation expansion method \cite{Yang2008,Yang2009}.

\begin{figure}
\begin{center}
\includegraphics[width=\columnwidth]{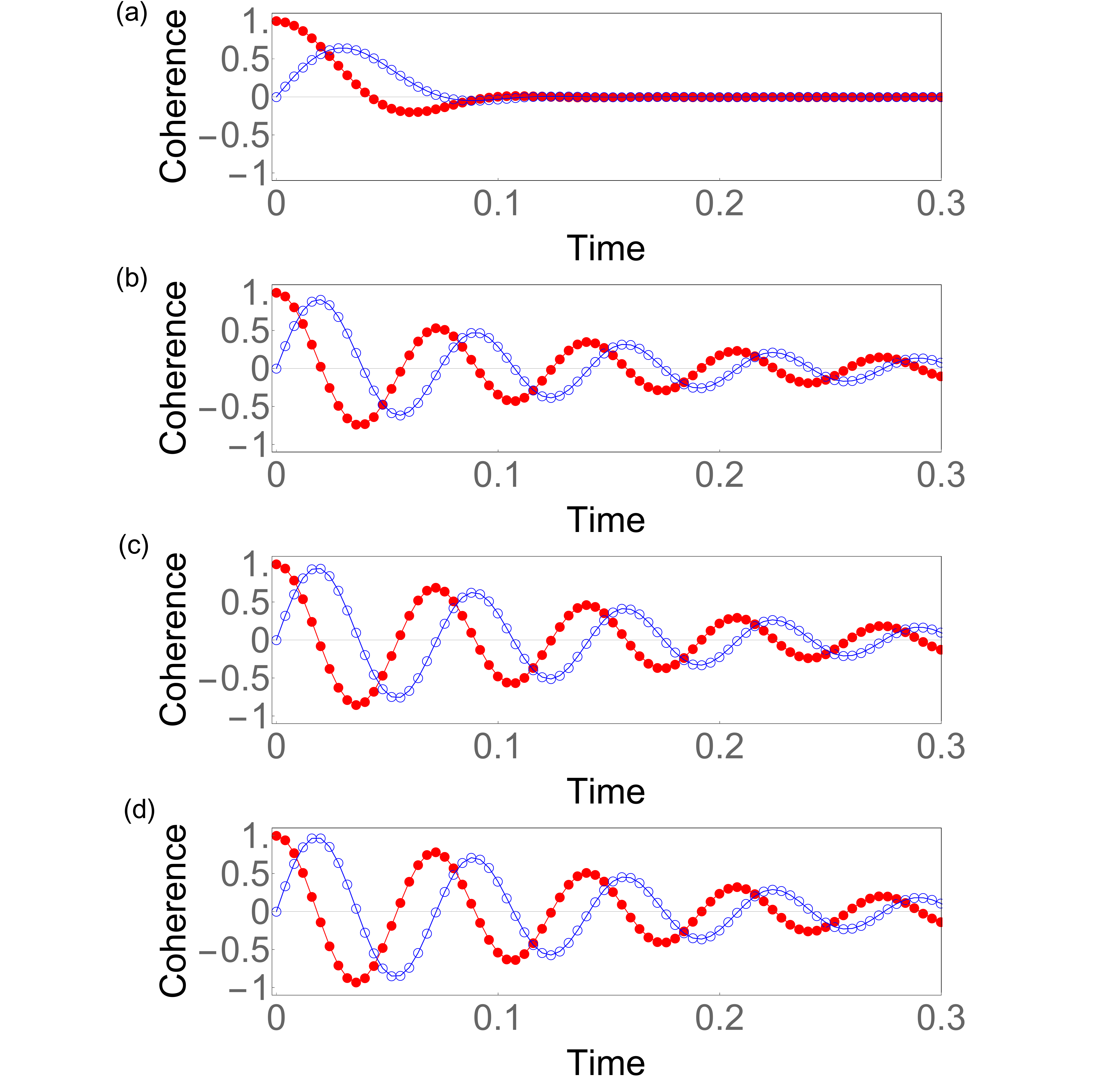}
\end{center}
\caption{(color online). Quantum decoherence of the probe spin which is coupled to the two-dimensional Ising model at inverse temperature $\beta=\beta_c$ and magnetic field $h=0.1J$ as a function of time for different sizes of rectangle with the same area $M\times N=100$. The red solid dot is the real part of the coherence $\langle S_x(t)\rangle$ and the blue circle is the imaginary part of the coherence $\langle S_y(t)\rangle$. (a). Spin decoherence for $2\times50$ rectangular lattices. (b). Spin decoherence for $4\times25$ rectangular lattices. (c). Spin decoherence for $5\times20$ rectangular lattices. (d). Spin decoherence for $10\times10$ rectangular lattices.}
\label{fig:epsart4}
\end{figure}

\begin{figure}
\begin{center}
\includegraphics[width=\columnwidth]{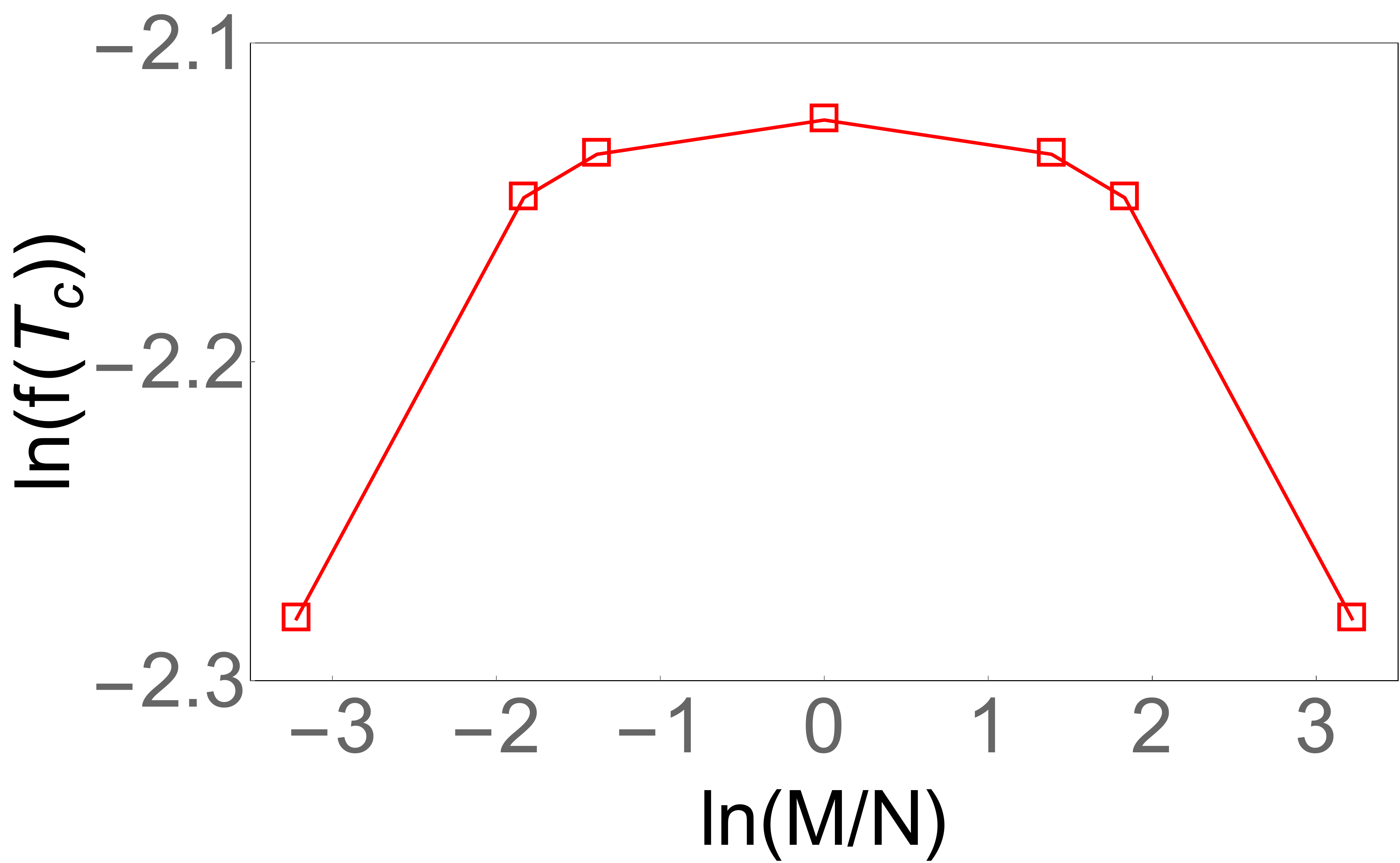}
\end{center}
\caption{(color online). The effect of elongation in a 2D lattice at conformal invariant critical point. The vertical axis shows the logarithm of the free energy at the critical point and the horizontal axis presents the logarithm of the ratio of the two edges of the rectangle, $x\equiv M/N$. All the data points have the same area $M\times N$. The squares denote the free energy reconstructed from the spin coherence data (Figure 4) by Equation \eqref{a}. }
\label{fig:epsart5}
\end{figure}

Figure 2 shows the spin coherence of the probe spin which is coupled to the two-dimensional Ising model at $\beta=\beta_c$ and $h=0.1J$ as a function of time for 2D rectangular lattice of different sizes $6\times50,7\times50,8\times50,9\times50,10\times50$ in (a)-(e), respectively. In Figure 2 (a)-(e), the red solid dot shows the real part of the spin coherence and the blue circle presents the imaginary part of the spin coherence. These data is used to obtain the free energy of the two-dimensional Ising model at zero magnetic field from Equation \eqref{a}. We uniformly pick up 394 data points of the spin coherence within one period $\eta t=\pi/2$ and make use of the Simpson-3/8 rule to numerically reconstruct the free energy at the critical point. The reconstructed free energy is shown in Figure 3 and labelled by black squares. From Equation \eqref{centralA}, we know that for a square lattice with $M\times N$ spins, if $M>>N$, then the free energy per spin is a linear function of $1/N^2$ with the slop being $-\frac{\pi c}{6}$. We then make a linear fit of the free energy as a function of $1/N^2$ and found that the slop is $-0.2636\pm0.0013$. From Equation \eqref{centralA}, we know that the central charge $c=0.2636\times 6/\pi\approx0.503\pm 0.003$. Thus the estimated central charge agrees with the exact solution $c=1/2$ perfectly.

We further test the effect of elongation in conformal field theory corresponding to the two-dimensional critical point. Figure 4 shows the spin coherence of the probe spin which is coupled to the 2D Ising model at $\beta=\beta_c$ and $h=0.1J$ as a function of time for 2D rectangular lattices with the same area, being $2\times50,4\times25,5\times20,10\times10$ in (a)-(d), respectively. In Figure 4 (a)-(d), the red solid dot shows the real part of the spin coherence and the blue circle presents the imaginary part of the spin coherence. We then make use of these data to construct the free energy of 2D Ising model at zero magnetic field from Equation \eqref{a}. We uniformly pick up 394 data points of the spin coherence within one period and make use of the Simpson-3/8 rule to numerically reconstruct the free energy at the critical point. The reconstructed free energy is shown in Figure 5 and labelled by squares. We show the logarithm of the free energy at the critical point as a function of the logarithm of the ratio of two edges of the rectangular lattice. First, one can see that the free energy is invariant under the modular transformation, interchanging of $M$ and $N$, i.e. $x\leftrightarrow1/x$. Second, the free energy of a 2D domain with a fixed area is maximum for square $x=1$ and decreases as $x$ becomes larger or smaller. This means there is a thermodynamic driving force for elongation of the a rectangular domain.

In summary, we show that the central charge of the conformal field theory corresponding to a critical point of a two-dimensional lattice models can be extracted from the quantum decoherence measurement of a probe spin which is coupled to the two-dimensional lattice models. Measuring the quantum coherence of a
single probe spin provides a new approach to studying the universality of two-dimensional interacting many-body systems.

\begin{acknowledgements}
This work was supported by the National Natural Science Foundation of China under Grant NO. 11604220 and the Startup Fund of Shenzhen University under Grant NO. 2016018.
\end{acknowledgements}


\begin{references}

\bibitem{Stanley1971}
H. E. Stanley, \emph{Introduction to Phase Transitions and Critical Phenomena} (Clarendon Press, Oxford, 1971).

\bibitem{Cardy1996}
J. L. Cardy, \emph{Scaling and Renormalization in Statistical Physics} (Cambridge University Press, Cambridge, 1996).


\bibitem{BPZ1984}
A. A. Belavin, A. M. Polyakov and A.B. Zamolodchikov, Infinite conformal symmetry in two-dimensional quantum field theory, Nucl. Phys. B \textbf{241}, 333 (1984).

\bibitem{conformalbook}
M. Henkel,\emph{Conformal Invariance and Critical Phenomena} (Springer-Verlag Berlin Heidelberg, 1999).

\bibitem{Virasoro1970}
M. A. Virasoro, Subsidiary Conditions and Ghosts in Dual-Resonance Models, Phys. Rev. D \textbf{1}, 2933 (1970).

\bibitem{Shenker1984}
D. Friedan, Z. Qiu and S. Shenker, Conformal Invariance, Unitarity, and Critical Exponents in Two Dimensions, Phys. Rev. Lett. \textbf{52}, 1575 (1984).


\bibitem{Kac1979}
V. G. Kac, in Group Theoretical Methods in Physics, edited by Beiglbock and A. Bohm, Lecture Notes in Physics
Vol. 94 (Springer-Verlag, New York, 1979), p. 441.


\bibitem{Cardy1985}
H. W. J. Bl\"{o}te, J. L. Cardy and M. P. Nightingale, Conformal Invariance, the Central Charge, and Universal Finite-Size
Amplitudes at Criticality, Phys. Rev. Lett. \textbf{56}, 742(1985).

\bibitem{Affleck1985}
I. Affleck, Universal Term in the Free Energy at a Critical Point and the Conformal Anomaly, Phys. Rev. Lett. \textbf{56}, 746(1985).



\bibitem{Cardy1988}
J. L. Cardy, Fields, Strings and Statistical Mechanics ed by E. Brhzin and J Zinn-Justin (Amsterdam, Noah-Holland, 1988) pp 169-246.


\bibitem{Torus1986}
C. Itzykson and J. B. Zuber, Two-dimensional conformal invariant theories on a torus, Nucl. Phys. B \textbf{275},580(1986).









\bibitem{Wei2012}
B. B. Wei and R. B. Liu, Lee-Yang zeros and critical times in decoherence of a probe spin coupled to a bath, Phys. Rev. Lett. \textbf{109}, 185701 (2012).

\bibitem{Wei2014}
B. B. Wei, S. W. Chen, H. C. Po and R. B. Liu, Phase transitions in the complex plane of a physical parameter, Sci. Rep. \textbf{4}, 5202 (2014).

\bibitem{Wei2015}
B. B. Wei, Z. F. Jiang and R. B. Liu, Thermodynamic holography, Sci. Rep. \textbf{5}, 15077 (2015).

\bibitem{Peng2015}
X. H. Peng, H. Zhou, B. B. Wei, J. Y. Cui, J. F. Du and R. B. Liu, Experimental observation of Lee-Yang zeros, Phys. Rev. Lett. \textbf{114}, 010601 (2015).

\bibitem{LYExp2015}
N. Ananikian and R. Kenna, Imaginary magnetic fields in the real world, Physics \textbf{8}, 2 (2015).

\bibitem{Wei2017}
B. B. Wei, Probing Yang-Lee edge singularity by central spin decoherence, New Journal of Physics (in press) (2017).

\bibitem{scexp1}
L. Childress \emph{et al}. Coherent Dynamics of Coupled Electron and Nuclear Spin Qubits in Diamond, Science \textbf{314}, 281 (2006).

\bibitem{scexp2}
R. Hanson, V.V. Dobrovitski, A. E. Feiguin, O. Gywat, and D. D. Awschalom, Coherent Dynamics of a Single Spin Interacting with an Adjustable Spin Bath, Science \textbf{320}, 352 (2008).


\bibitem{scexp3}
H. Bluhm \emph{et al}. Dephasing time of GaAs electron-spin qubits coupled to a nuclear bath exceeding 200 $\mu$s, Nature Phys. \textbf{7}, 109 (2010).

\bibitem{scexp4}
Y. Li \emph{et al}. Intrinsic Spin Fluctuations Reveal the Dynamical Response Function of Holes Coupled to Nuclear Spin Baths in (In,Ga)As Quantum Dots, Phys. Rev.
Lett. \textbf{108}, 186603 (2012).

\bibitem{scexp5}
N. Zhao \emph{et al}. Sensing single remote nuclear spins, Nature Nanotech. \textbf{7}, 657 (2012).


\bibitem{Onsager1944}
L. Onsager, Crystal statistics. I. A Two-dimensional model with an order-disorder
transition, Phys. Rev. \textbf{65}, 117(1944).

\bibitem{Kramers1941}
H. A. Kramers and G. H. Wannier, Statistics of the two-dimensional Ferromagnet.
Part I., Phys. Rev. \textbf{60}, 252(1941).


\bibitem{Wu1973}
 B. McCoy, T. T. Wu, The two-dimensional Ising model (Harvard University Press, Cambridge, 1973).

\bibitem{Schultz1964}
T. D. Schultz, D. C. Mattis and E. H. Lieb, Two-dimensional Ising model as a
soluble problem of many fermions, Rev. Mod. Phys.
\textbf{36}, 856 (1964).

\bibitem{Reichl2016}
L. E. Reichl, A Modern Course in Statistical Physics, 4th ed. (Wiley, New York, 2016).

\bibitem{Yang2008}
W. Yang and R. B. Liu,  Quantum many-body theory of qubit decoherence in a finite-size spin bath, Phys. Rev. B \textbf{78}, 085315 (2008).

\bibitem{Yang2009}
W. Yang and R. B. Liu, Quantum many-body theory of qubit decoherence in a finite-size spin bath. II. Ensemble dynamics, Phys. Rev. B \textbf{79}, 115320 (2009).




\end{references}
\end{document}